# Effects of external mechanical loading on stress generation during lithiation in Li-ion battery electrodes

Wenbin Zhou[*]

*Department of Mechanical Engineering, Imperial College London, London SW7 2AZ, UK*

**Abstract**

Li-ion batteries are ineluctably subjected to external mechanical loading or stress gradient. Such stress can be induced in battery electrode during fabrication and under normal operation. In this paper, we develop a model for stresses generated during lithiation in the thin plate electrode considering the effects of external mechanical loading. It is found that diffusion-induced stresses are asymmetrically distributed through the thickness of plate due to the coupling effects of asymmetrically distributed external mechanical stress. At the very early stage during Li-ions insertion, the effects of the external mechanical loading is quite limited and unobvious. With the diffusion time increasing, the external mechanical loading exerts a significant influence on the evolution of stresses generated in the electrode. External compressed electrode is inclined to increase the value of stresses generated during lithiation, while external tensed electrode tends to decrease the value of stresses, and as the diffusion time increases, the effects of the external mechanical loading on the stresses generated during lithiation become more obvious.

[*]Corresponding author. Email address: w.zhou15@imperial.ac.uk, wbzhou@pku.edu.cn (W. Zhou)

## 1. Introduction

Li-ion batteries have been widely used in mini-type electric instruments such as cell phones and other small portable electronic devices due to their high specific capacity, light weight, and no memory effects. However, their further applications such as in the fields of electromotive vehicle, large-scale energy storage and aerospace power supply are limited by the poor cycle of life and the capacity loss of themselves. The main reason is due to the deformation and fracture of the electrode caused by the stress generated in electrodes during cycling [1-4], which can result in electrical disconnects that render electrode active materials incapable of storing Li-ions. One of the critical challenges in Li-ion battery studies is the internal stress field distribution produced in the process of charging and discharging.

Much work has been devoted to studying the stresses resulting from cycling of Li-ion batteries due to the intercalation and deintercalation of Li-ions [5-11], however, studies for effects of external mechanical loading on stress generation during lithiation in Li-ion battery electrodes are less. Actually, Li-ion batteries are ineluctably subjected to external mechanical loading or stress gradient during standard usage or storage. Such stress can arise during fabrication of battery materials, which usually involves compression of the electrode to control its porosity [12-14]. Also, such stress can be induced in electrode materials under normal operation, such as batteries operating in high pressure environments, electric/hybrid vehicles or flexible applications [15]. In many cases, the influence of external stresses on the diffusion process and the stress generation during lithiation in Li-ion battery electrodes is so significant that it cannot be neglected. Some researchers have studied the effects of electrode compression on capacity and efficiency of various electrode materials. Novak *et al.* [16] and Gnanaraj *et al.* [17] found both the capacity and first cycle efficiency of batteries had a decrease when subjected to high levels of compression. They explained that it was caused by electrode particle fracture for the largest pressures as well as transport limitations within the liquid path. By studying low-loading electrodes prepared with a natural graphite from Superior Graphite and compressed at a range of moderate pressures, Shim *et al.* found both the reversible capacity and irreversible capacity loss (ICL) had a decrease with increases in pressure [18]. They concluded that the decreased reversible capacity was due to increased stresses generated within the graphite electrode, which also slowed down in Li-ions diffusion process [19]. However, much of these work were conducted using electrochemical characterization. Theoretical studies for the effects of external mechanical loading on stress generation during lithiation in electrodes which are quite necessary for Li-ion batteries have been rare.

In this work, we develop a model for stress generation during lithiation in Li-ion battery electrodes with planar geometries coupled with the effects of external mechanical loading. To begin with, two basic approximations for material properties are introduced as follows [9,10]. Isotropic elasticity analysis of infinitesimal deformation is carried out, and large deformation related to plasticity is out of the scope of this research [20]. Mechanical properties and the diffusion coefficient are independent of Li-ion concentration, which means that Young's modulus, Poisson's ratio and the diffusion coefficient are constants. Yang *et al*. [21] studied the effects of composition-dependent modulus, finite concentration and boundary constraint on Li-ion diffusion and stresses in a bilayer Cu-coated Si nano-anode. This is an important work to show that mechanical properties of electrode materials will change and have an effect on diffusion-induced stresses during the lithiation/delithiation process. Since we focus on studying the effects of external mechanical loading, changes in material properties during the lithiation/delithiation process still need to be further considered by establishing more sophisticated model.

## 2. Basic Theory

### 2.1. Diffusion equation

We model the insertion and extraction of Li-ions as a diffusion process. The species flux can be defined as [22]

$$J = -Mc\nabla\mu \quad (1)$$

where $M$ is the mobility of Li-ions, $c$ is the Li-ion concentration and $\mu$ is the chemical potential, which is given by

$$\mu = \mu_0 + RT\ln c - \Omega\sigma_h \quad (2)$$

where $\mu_0$ is the initial chemical potential and is assumed to be a constant. $R$ is a gas constant, $T$ is absolute temperature, $\Omega$ is the partial molar volume of the Li-ion, $\sigma_h$ is the hydrostatic stress. Since atomic diffusion in solids is much slower than elastic deformation, mechanical equilibrium is established much faster than that of diffusion. Mechanical equilibrium is, therefore, treated as a static equilibrium problem, thus the linear elasticity theory is applicable for the coupling of diffusion-induced stresses and external stresses. The hydrostatic stress $\sigma_h$ can be expressed as

$$\sigma_h = \frac{1}{3}\sum_{i=x,y,z}(\sigma_{di} + \sigma_{ei}) \quad (3)$$

where $\sigma_{di}$ is the diffusion-induced stress due to the concentration gradient and $\sigma_{ei}$ denotes the external mechanical loading induced stress, respectively.

Substituting Eq. (2) into Eq. (1), the species flux can be expressed as

$$J = -MRT(1 - \frac{\Omega c}{RT}\frac{\partial \sigma_h}{\partial c})\nabla c = -D_{eff}\nabla c \quad (4)$$

where $D_{eff}$ is the effective diffusion coefficient in a stressed isotropic solid. Substituting Eq. (3) into Eq. (4), $D_{eff}$ can be given by

$$D_{eff} = D(1 - \frac{\Omega c}{RT}\frac{\partial \sigma_h}{\partial c}) = D\left\{1 - \frac{\Omega c}{3RT}\frac{\partial\left[\sum_{i=x,y,z}(\sigma_{di} + \sigma_{ei})\right]}{\partial c}\right\} \quad (5)$$

where $D = MRT$ is the diffusion coefficient in a stress-free isotropic solid. Combining Eq. (4) with the mass conservation equation $\frac{\partial c}{\partial t} + \nabla \cdot J = 0$, we obtain the diffusion equation as follows

$$\frac{\partial c}{\partial t} = D\left\{\nabla^2 c - \frac{\Omega c}{3RT}\nabla^2\left[\sum_{i=x,y,z}(\sigma_{di} + \sigma_{ei})\right] - \frac{\Omega \nabla c}{3RT}[\nabla\left[\sum_{i=x,y,z}(\sigma_{di} + \sigma_{ei})\right]]\right\} \quad (6)$$

**2.2. Diffusion-induced stresses**

Consider an electrode plate of thickness *l* subjected to a constant uniform charging current density on both of its side faces [2, 23], as shown in **Fig. 1**. The electrode plate is considered to be an isotropic linear elastic solid and is mainly composed of the active particles. The effect of the electrolyte is neglected though it can be included using a more sophisticated model such as the work conducted by Zhang *et al*. [24] and Renganathan *et al*. [25], and the aim of our work is to study the effects of external mechanical loading. Analogous to thermal stresses [26], for a given concentration profile, the nonzero diffusion-induced stress components due to the insertion of solute atoms into host are only $\sigma_{dy}$ and $\sigma_{dz}$, which are two equal transverse stresses and can be given by

$$\sigma_{dy} = \sigma_{dz} = \frac{E\Omega}{3(1-\nu)}(\frac{1}{l}\int_0^l cdx - c) + \frac{2(2x-l)}{l^3(1-\nu)}\int_0^l E\Omega c(x - \frac{l}{2})dx \quad (7)$$

where *E* and $\nu$ are Young's modulus and Poisson's ratio, respectively.

In general, the thin electrode plate is subjected to a unidirectional gradient stress field due to the mechanical loading combination of bending $M_z$ and tension $F_y$. As illustrated in **Fig. 1**, the unidirectional gradient stress field can be expressed as

$$\sigma_{ex} = \sigma_{ez} = 0, \quad \sigma_{ey} = p_0 x + a_0 \quad (8)$$

where $p_0$ denotes the stress gradient due to mechanical bending and $a_0$ is the tensile stress. It should be noted that **Fig.1** only shows the case of $p_0 > 0$, in fact, $p_0$ can be negative and zero and are also studied.

Substituting the diffusion-induced stress Eq. (7) and external loading induced stress Eq. (8) into Eq. (6) yields

$$\frac{\partial c}{\partial t} = D\left\{ (1+\theta c)\nabla^2 c + \theta(\nabla c)^2 - \frac{12\theta}{l^3}\nabla c \int_0^l (x-\frac{l}{2})c\,dx - \frac{\Omega p_0}{3RT}\nabla c \right\} \quad (9)$$

where $\theta = \dfrac{2\Omega^2 E}{9RT(1-\nu)}$. Hence, the effective diffusion coefficient $D_{eff}$ in Eq. (5) becomes

$$D_{eff} = D\left\{ 1+\theta c - \frac{12\theta c}{l^3 \nabla c}\int_0^l (x-\frac{l}{2})c\,dx - \frac{\Omega c p_0}{3RT\nabla c} \right\} = D(1+\beta c) \quad (10)$$

where $\beta = 1+\theta - \dfrac{12\theta}{l^3 \nabla c}\int_0^l (x-\dfrac{l}{2})c\,dx - \dfrac{\Omega p_0}{3RT\nabla c}$. It is a positive constant and reflects the degree of the effects of both the concentration gradient and external stress gradient on diffusion flux.

Two operations reflect the insertion and extraction of Li-ions: surface Li-ion concentration keeps the maximum concentration and surface Li-ion flux remains a constant flux [7]. In this work, we assume that the surface Li-ion flux is uniform, namely galvanostatic operation, and the initial and boundary conditions are given by

$$\begin{cases} c(x,0) = c_0, & t=0 \\ D_{eff}\nabla c = J_b, & x=0 \\ D_{eff}\nabla c = J_t, & x=l \end{cases} \quad (11)$$

where $J_b$, $J_t$ are the Li-ion flux at the bottom and top surface, respectively. To perform numerical calculations conveniently, the following dimensionless variables for thickness $\hat{x}$, concentration $\hat{c}$, time $\hat{t}$, parameter $\hat{\theta}$, current density $\hat{j}$, stress $\hat{\sigma}$, and external mechanical stress gradient $P$, are also introduced:

$$\hat{x} = \frac{x}{l},\ \hat{c} = \frac{c}{c_{max}},\ \hat{t} = \frac{tD}{l^2},\ \hat{\theta} = \theta c_{max},\ \hat{j} = \frac{Jl}{Dc_{max}},\ \hat{\sigma} = \frac{3(1-\nu)}{E\Omega c_{max}}\sigma,\ P = \frac{\Omega l}{3RT}p_0 \quad (12)$$

Finite difference method is adopted here to solve the non-linear Eq. (9) combined with definite conditions in Eq.

(11), where the time derivative and space derivative of the Li-ion concentration both need to be discretized. The time derivative and space derivative are calculated by the forward difference and central difference, respectively. The Simpson's rule is used to calculate the integral. The dimensionless time increment is set as $\Delta \hat{t}=0.001$. The dimensionless thickness of the plate is divided into 200 equal parts, i.e. $\Delta \hat{x}=0.005$. Once the diffusion equation is solved, Li-ion concentration is substituted into Eq. (7) for stresses generated in the electrode.

## 3. Results and Discussion

We take the $Mn_2O_4$ system as an example, whose material properties are listed in Table 1 [10]. For simplicity, the surface flux of Li-ions at the bottom surface $J_b$ is assumed to be identical with that of the top surface $J_t$. As an example, we take the dimensionless current density at the bottom and top surface as $J_b l / D c_{max} = 0.1$, and $\theta c_{max} = 0.4$. We state that different fluxes could influence the stress profiles because of different Li-ion concentrations at the two surfaces. Eq. (11) can be applied to the Li-ion diffusion during charging and discharging: $c_0 = 0$, $J_b = -J_{bs}$, and $J_t = J_{ts}$ for the process of Li-ions insertion, while $c_0 = c_{max}$, $J_b = J_{bs}$, and $J_t = -J_{ts}$ for Li-ions deinsertion.

**Fig. 2** shows the results of insertion in the absence of external mechanical stress gradient: (a) normalized concentration profile through the thickness of the plate, (b) normalized stress profile at different locations and (c) normalized stress over the surface at different charging times. At the early period during Li-ions insertion in **Fig. 2(a)**, Li-ions transfer across the two surfaces first and gradually move towards the centre, showing that Li-ion concentration is much higher around the surface compared with the region close to the centre. Therefore, surface region experiences larger volume changes, which causes tensile stress in the centre and compressive stress in the two surface regions as shown in **Fig. 2(b)**. Also, both the concentration and stress are symmetrically distributed through the thickness of the plate in the absence of external mechanical stress gradient, and the maximum stress is always located at the entry surface for a given time. **Fig. 2(c)** illustrates the evolution of normalized stress at the entry surface versus time, we can see that the magnitude of this stress increases to its peak with increasing time and then gradually decreases as the charging process proceeds. This is because stress is relevant to the concentration profile of Li-ions inserted into the plate, at the initial charging stage, the concentration profile is nonuniform, as more and more Li-ions insert into the plate, the concentration profile becomes more and more uniform, resulting in gradually decreasing stress.

To study the influence of external mechanical stress, we set five different $P$: 0, -0.5, -1, 0.5 and 1. $P = \frac{\Omega l}{3RT} p_0$ is the dimensionless form of $p_0$, which is given by Eq. (8) and denotes the stress gradient. Here the negative stress gradient $P$ means applying external compressive loading to the electrode, while positive $P$ means applying external tensile loading. **Fig. 3** shows the results of insertion with different negative external mechanical stress gradients: (a) normalized concentration profile at the stage of dimensionless time $Dt/l^2 = 0.16$, (b) normalized stress profile at the early stage of dimensionless time $Dt/l^2 = 0.032$ during Li-ions insertion and (c) normalized stress profile at the later stage of dimensionless time $Dt/l^2 = 0.07$ during Li-ions insertion. From **Figs. 3(b)** and **3(c)**, we can see that stress profile through the thickness of the plate is asymmetric due to the coupling effects of asymmetrically distributed external mechanical stress. At the early stage during Li-ions insertion ($Dt/l^2 < 0.05$), the effects of the external mechanical loading is quite limited and unobvious. With the charging time increasing, the external stress gradient exerts a significant influence on the evolution of stresses. Such an influence will increase with increasing external stress gradient, and larger external negative stress gradient tends to increase the value of stresses generated in the electrode compared with that of lower external negative stress gradient as time increases. This result is in accordance with the previous experiments [16-18] that compressed electrode has a decrease in capacity. It also implies that when the external negative stress gradient is applied and becomes larger, the charging at the entry surface will be slower. The external negative stress gradient will impede the diffusion of solute atoms in this case. Consequently, the value of the concentration decreases and thus leads to more unevenly distributed concentration, as shown in **Fig. 3(a)**.

The results of insertion with different positive stress gradients are illustrated in **Fig. 4**: (a) normalized concentration profile at the period of dimensionless time $Dt/l^2 = 0.12$, (b) normalized stress profile at the early period of dimensionless time $Dt/l^2 = 0.032$ during Li-ions insertion and (c) normalized stress profile at the later period of dimensionless time $Dt/l^2 = 0.07$ during Li-ions insertion. In contrast to **Fig. 3**, here external tensile loading is applied to the electrode. As expected, the external positive stress gradient will accelerate the solute penetration and thus leads to more uniform concentration profile, seen in **Fig. 4(a)**. Therefore, the external tensile stress can be employed to modify diffusion barriers and help to decrease strains/stresses originating from unevenly distributed concentration. Yen *et al.* found tensile mechanical stress will enlarge atom spacing of silicon and thus enhance the oxidation rate [27]. Moreover, Sanchez *et al.* showed that an external tensile stress of 2GPa decreased diffusion barriers by about 9%, improving diffusion rates by about 30% at room temperature [28]. As shown in **Figs. 4(b)** and **4(c)**, the external positive stress gradient tends to decrease the value of stresses generated in the electrode, and as the charging time increases, the effects of external stress gradient become more obvious. The deformation and fracture of the electrode caused by the excessive internal stress during cycling can result in electrical disconnects, which renders electrode active material incapable of storing lithium-ion and lows the utilization of the active material. From **Figs. 3(c)** and **4(c)**, at the same diffusion time $Dt/l^2 = 0.07$ during Li-ions insertion, the maximum tensile stress ($\hat{\sigma}$ =0.03) of electrode subjected to positive external stress gradient $P$=1 are reduced by as much as 86% compared to electrode subjected to negative external stress gradient $P$=-1 ($\hat{\sigma}$ =0.22). Therefore, tensed electrode is superior in fracture resistance, and thus the utilization of the active material should be significantly increased because of the decrease of the maximum tensile stresses generated during cycling. By the results mentioned above, diffusion-induced stresses can be tailored by the external stress gradient in order to be kept below material strengths and avoid mechanical fracture. To quantify the effects of the non-uniform distribution of the stresses on the utilization of the active material, more data such as the critical fracture strength of electrode materials and fracture energy are needed from a range of further experiments.

## 4. Conclusions

In summary, we develop a model for diffusion-induced stresses of the thin plate electrode and consider the effects of external mechanical loading. The results show that stress profile through the thickness of the plate is asymmetric due to the coupling effects of asymmetrically distributed external mechanical stress. At the early stage during Li-ions insertion, no significant change is observed for different external stress gradients. With the increase of the diffusion time, the external stress gradient exerts significant effects on the evolution of stresses. Such effects become more obvious with increasing the external stress gradient, larger external negative stress gradient leads to greater value of stresses as time increases, while larger external positive stress gradient tends to decrease the value of stresses generated in the electrode, and the effects become more obvious as time increases.

**Tables**

Table 1　Material parameters of $Mn_2O_4$

| Name | Symbol | Value |
|---|---|---|
| Diffusion coefficient | $D$/m$^2$ s$^{-1}$ | $7.08\times10^{-15}$ |
| Young's modulus | $E$/GPa | 10 |
| Poisson's ratio | $\nu$ | 0.3 |
| Partial molar volume | $\Omega$/dm$^3$ mol$^{-1}$ | $3.497\times10^{-3}$ |
| Max Li-ion concentration | $c_{max}$/mol dm$^{-3}$ | 22.9 |

**Figures and Figure Captions**

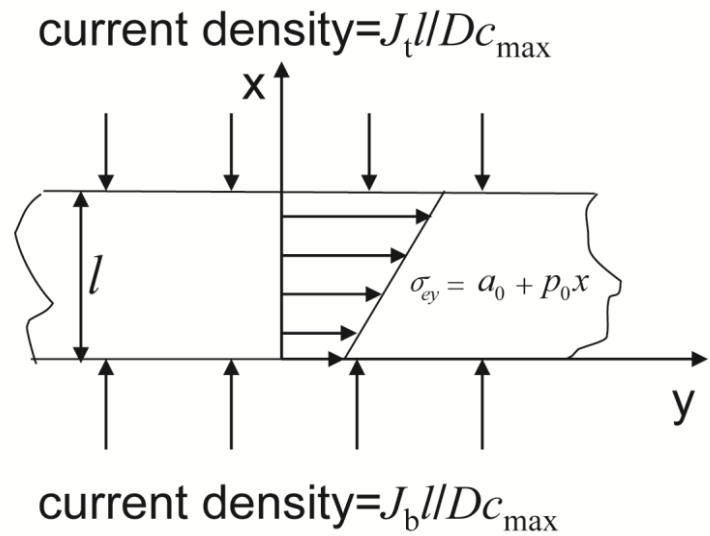

**Fig.1** Sketch of a plate electrode under galvanostatic charging, namely a uniform current density on surfaces.

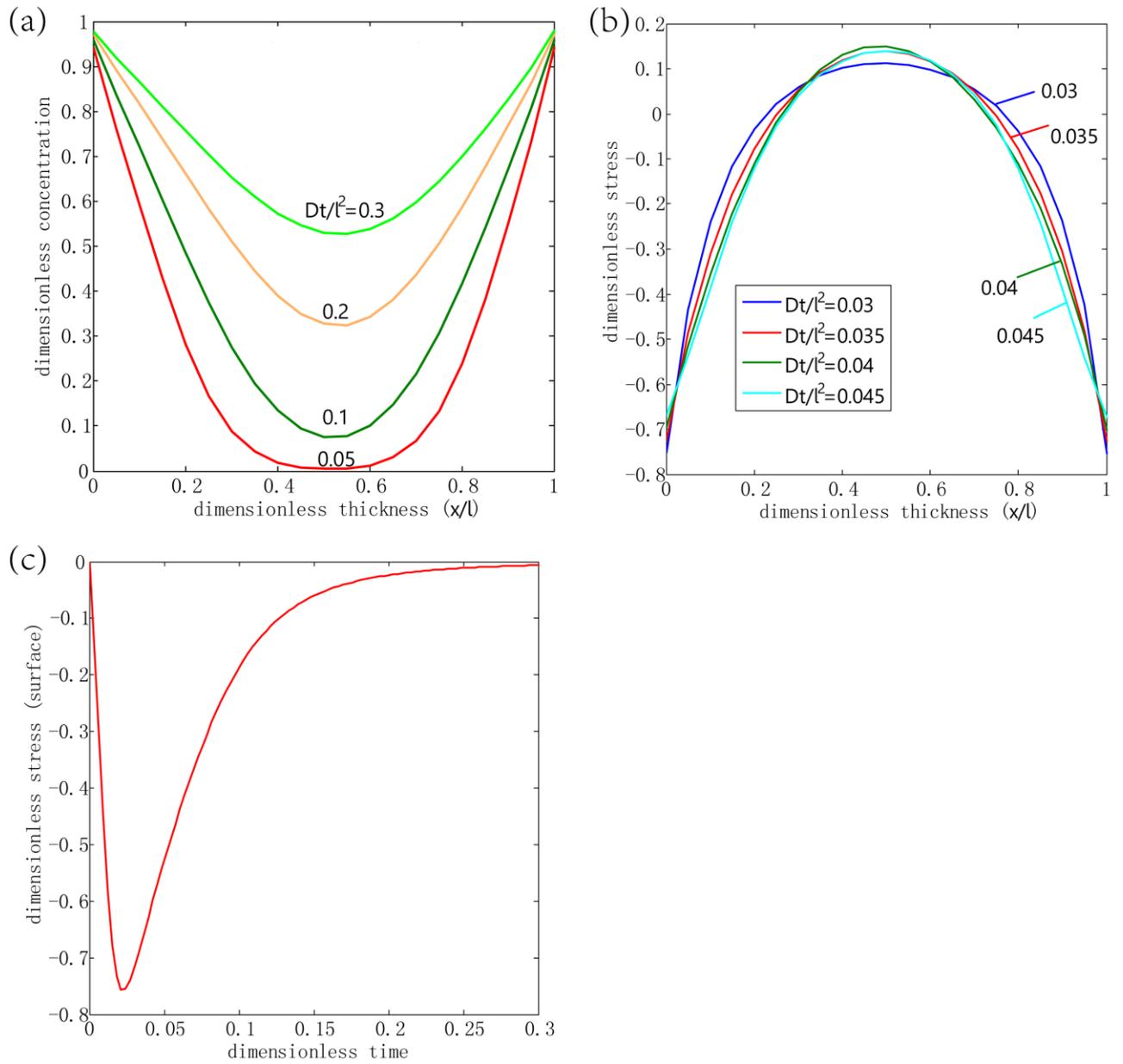

**Fig.2    (a) Evolution of the normalized concentration profile, (b) Evolution of the normalized stress profile and (c) Evolution of the normalized stress over the surface.**

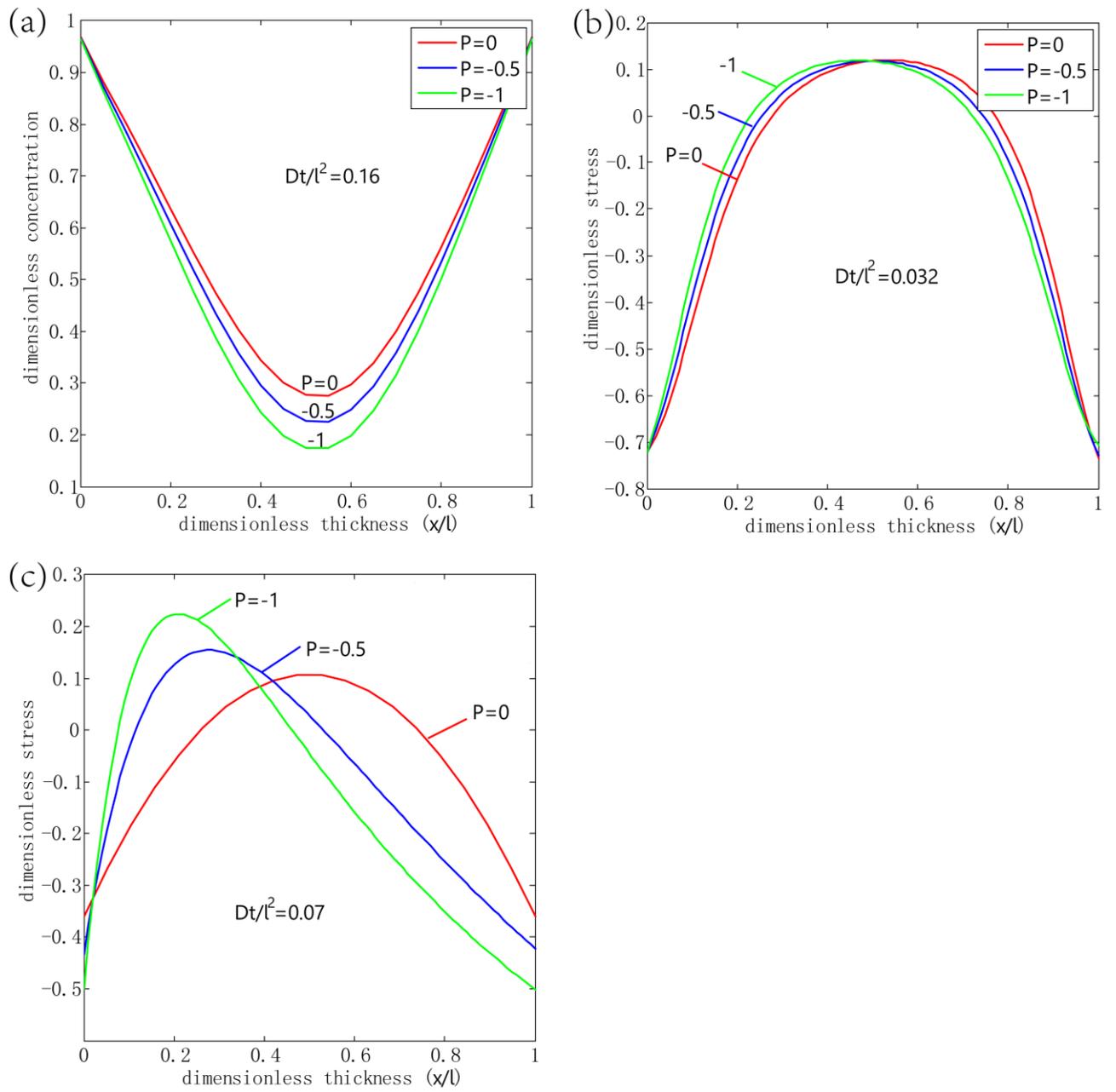

**Fig.3** The effects of the negative stress gradient on: (a) concentration profile at the diffusion time $Dt/l^2 = 0.16$, (b) stress profile at $Dt/l^2 = 0.032$ and (c) stress profile at $Dt/l^2 = 0.07$.

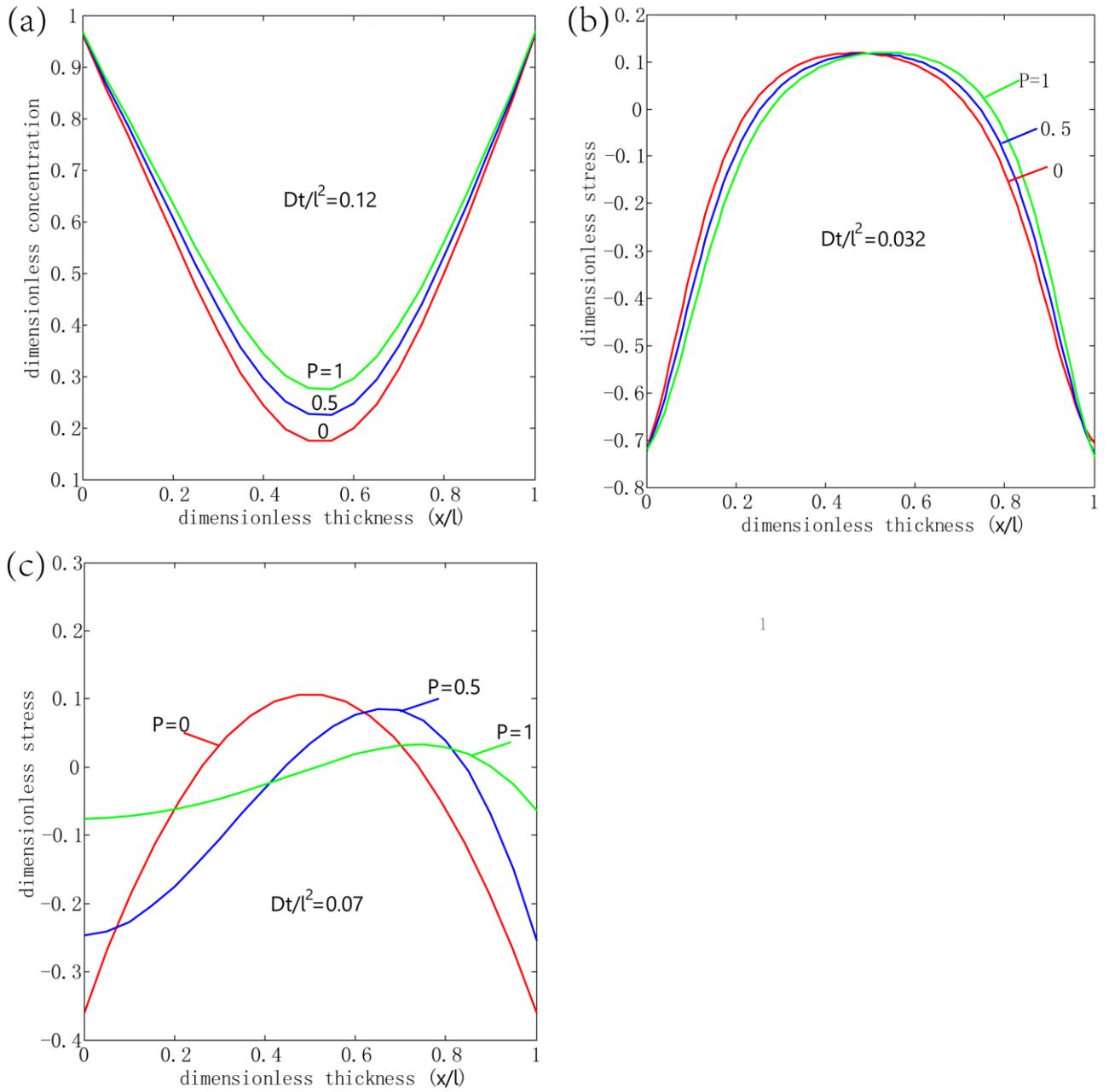

**Fig.4** The effects of the positive stress gradient on: (a) concentration profile at the diffusion time $Dt/l^2 = 0.12$, (b) stress profile at $Dt/l^2 = 0.032$ and (c) stress profile at $Dt/l^2 = 0.07$.